\documentclass[5p,times]{elsarticle}
\usepackage{lineno,hyperref,amsmath,comment,float,pdfpages,graphicx,}
\modulolinenumbers[50]

\journal{International Journal of Heat and Fluid Flow}

\usepackage{numcompress}\biboptions{authoryear}
\begin{document}

\begin{frontmatter}

\title{Two-phase flow simulations at $0-4^{o}$ inclination in an eccentric annulus}
%%\tnotetext[mytitlenote]{Fully documented templates are available in the elsarticle package on \href{http://www.ctan.org/tex-archive/macros/latex/contrib/elsarticle}{CTAN}.}

%% Group authors per affiliation:
\author[mymainaddress]{C. Friedemann \corref{mycorrespondingauthor}}
\cortext[mycorrespondingauthor]{Corresponding author}
\ead{chrisjfr@student.matnat.uio.no}

%% or include affiliations in footnotes:
\author[mymainaddress]{M.Mortensen}
%%\ead[url]{www.elsevier.com}

\author[mysecondaryaddress]{J. Nossen }

\address[mymainaddress]{Department of Mathematics. University of Oslo, Moltke Moes vei 35, 0851 Oslo, Norway}
\address[mysecondaryaddress]{Institute for Energy Technology, 2007 Kjeller, Norway}

\begin{abstract}
Multiphase flow simulations  were run in an eccentric annulus. The dimensions of the annulus were  0.1 and 0.05 m for the outer and inner cylinders, respectively, and the mixture velocities were varied between 1.2 and 4.2 m/s. The simulations were compared with fully eccentric and completely concentric experiments conducted at the Institute for Energy Technology in Norway. The purpose of this paper is to explore the effect of the holdup fraction and interior pipe's position on the pressure gradient and flow regime. The comparisons indicate that moving the pipe from an entirely eccentric to partially eccentric configuration has a drastic impact on the pressure gradient. In all cases where the inner pipe was changed from a completely eccentric geometry to a less eccentric configuration, we notice an increase of 48-303 \% of the mean pressure gradient. 
Comparatively, the cases where the pipe was moved from a concentric to a more eccentric configuration result in less drastic pressure gradient changes. 2 cases were within 22 \% of the experimental results for mean, maximum, and minimum pressure gradient, while the last two cases exceeded the minimum and mean pressure gradients by 25-250 \%, respectively.
We rarely observed a change of  flow regime as an effect of moving the inner pipe;  2 out of the 8 horizontal cases indicate either a transition from wavy flow to slug flow or significantly larger waves. The most prominent and frequent discrepancies identified were altered slug and wave frequencies. Through the simulations, we notice that there is an increased pressure gradient accompanying an increased holdup fraction when the phase-averaged velocities were the same. Corresponding to a fractional holdup increase of 0.177, 0.244, 0.063, and 0.073, the increase in simulated pressure gradient for each case of the same mixture flow rate and mesh density was 80, 300, 614 and 367 Pa/m respectively or 116, 244, 61.5 and 25 \%.  

\end{abstract}


\end{frontmatter}

\linenumbers

\section{Introduction}
Modern-day multiphase flow was spurred on by the discovery of oil and gas and the worldwide dependence on fossil fuels to power not only our cars and planes but countless other applications of everyday life.

 The topic of this paper, which is multiphase flow within an annulus, has direct applications to oil and gas extraction. However, the intentional usage of the annulus geometry is prevalent in several industries beyond petroleum.  One example is nuclear reactors.
 \cite{Sato} studied loss of coolant during an accident, which is highly applicable to the oil and gas industry, because of the similarities to a leak or rupture along a petroleum pipeline. The conducted studies go past environmental aspects, and there are several papers on the utilization of annular fuel rods concerning internal and external cooling, for example, \cite{Deokule} and \cite{ Blinkov}. There are also more fundamental flow studies, such as the velocity distribution within the annular mixing chamber \citep{Sun}. 

Multiphase pipe flow is a rigorously studied subject within the field of petroleum engineering. It is well known that multiphase flows and flow regimes are highly dependent on fluid composition, flow rates, and pipe inclination. Inclined pipe flow is less studied  than horizontal and vertical configurations. However, the literature covers topics such as slug frequency \citep{Hernandez, Hout, Schulkes}, holdup profile \citep{Beggs, Bonnecaze}, pressure drop \citep{Strazza, Salem, Ilic, Ghajar}, mechanical losses \citep{Liu} and flow regime \citep{Archibong, Oddie}.

Although we frequently find these topics in the body of work related to multiphase flow, the subjects are rare when studying the annulus configuration specifically. Although the annulus configuration is a well-known problem, the existing literature on the topic is not as extensive nor as rigorous as it is for a conventional pipe geometry. Therefore, this paper will revolve around the aspect of eccentricity of the annulus and its potential effect on the flow regime, holdup pattern, and pressure gradient.

 One of the earliest works applicable to the problem at hand presented in this paper was the modeling of frictional pressure drop and was performed in the late 1940s by Lockhart and Martinelli, a model which was later expanded by \cite{Chen}. The original model was based on correlations and was the result of a study into separated two-phase flows. Correlation models are prone to errors when the case falls outside the original scope of the measurements used to develop the model, yet may produce reliable results when used appropriately.
 
 Some of the earliest annulus studies originated in the 1960s as \cite{Denton} completed his Master's thesis on the topic of turbulent flow in concentric and eccentric annuli. During the same period, \cite{Michiyoshi} studied fully turbulent flow in a concentric annulus, while \cite{Vaughn} and \cite{Wein} both studied non-Newtonian fluids in annuli. Although most of the early publications related to the annulus configuration concentrated on single-phase flow, it signaled the beginning of an emerging field of study.  
 
By the mid-1970s, there was significant progress in terms of prediction and modeling of flow regimes. \cite{Taitel} worked on flow regime transitions for two-phase gas-liquid flow, and presented a generalized flow regime map, while \cite{Hanks} studied the specifics of laminar flow stability in a concentric annulus. However, as was the case in the previous decade, the 1970s saw consistent work on single-phase flows, and the majority of the efforts focused on the use or development of correlations. 

The 1980s continued in the same trend as the preceding decade; among the publications were further works mapping out annulus flow regimes  \citep{Kelessidis, Caetano}. Also, new correlations and models for flow behavior in different configurations, including annulus, were developed \citep{Hoyland,Dukler, Taitel2}. In addition, \cite{Kelessidis2} studied the motion of large gas bubbles passing through the liquid in vertical annuli. 

During the late 1990s, \cite{Iyer} conducted simulations of buoyancy induced flow in an annulus, outlining the effect of between 1 and 4 perturbations of the wall within the domain. Iyer's work on computational fluid dynamics related to the annulus configuration was one of the first publications which utilized CFD to study the annulus configuration. Simultaneously, \cite{Buryuk} studied the theoretical heat transfer in laminar flow within a concentric annulus. Several researchers conducted experimental studies of multiphase flow behavior. \cite{Harvel} studied the different flow regimes in a vertical annulus by using optical techniques such as X-ray and tomography. 
\cite{Escudier} studied non-Newtonian fluids in a concentric annulus while concurrent studies on void fraction \citep{Hasan}, as well as the rise velocity of Taylor bubbles \citep{Das, Hills} were published.

By the 2000s, technology had finally advanced to the point where CFD simulations became noticeable in the literature with regards to annulus flow. Simulations were published focused on turbulent flow and heat transfer \citep{Nikitin} as well as natural convection \citep{Adachi, Mizushima, Yoo, Yu}. Along with a new influx of simulation-based studies concerning the annulus configuration, there was a significant uptick in published experimental research. Asymmetric phase distributions \citep{Das2}, flow structures in a vertical annulus \citep{Hibiki, Ozar}, flow pattern and pressure drop in an inclined annular channel \citep{Wongwises} and convection in a vertical eccentric annulus \citep{Hosseini} were among the published works.

Even though annulus flow has been studied since the early 1960s, the recent increase in interest for the annulus configuration is natural because of its prevalence in industry combined with technological advances.
 Because the annulus configuration is frequently present in industries with potential for significant adverse environmental impacts, such as petroleum and nuclear, it is logical that the topic is of lasting interest to engineers and researchers. It is imperative to better understand not just the statistics, but also the physics related to multiphase flow in annuli to help mitigate and prevent environmentally damaging accidents.

To better understand and predict the behavior of multiphase flow within an annulus, we compare simulations performed in OpenFOAM using the Volume of Fluid type solver interFoam with experimental data from Institute for Energy Technology (IFE) in Norway. The experiments are conducted in a fully concentric or eccentric configuration. We use the experimental data such as averaged phase velocities and holdup fraction as initial conditions; however, we alter the domain utilized in the simulations to study the potential effect of the interior pipe's location. Additionally, we pair each simulation with another simulation of the same phase-averaged velocity but different holdup fraction.
 The simulation pairings allow us to study the effect of the holdup fraction itself on the resultant flow regime and pressure behavior.

\section{Geometry and mesh}
The annulus geometry is defined by the enclosed inner cylinder and its location compared to the outer cylinder. Their relative locations define the parameter known as eccentricity, which is one of the factors along with holdup fraction, which we will investigate to what degree affects the flow.

\begin{figure}[H]
\begin{center}
{\includegraphics[width = 9.0cm]{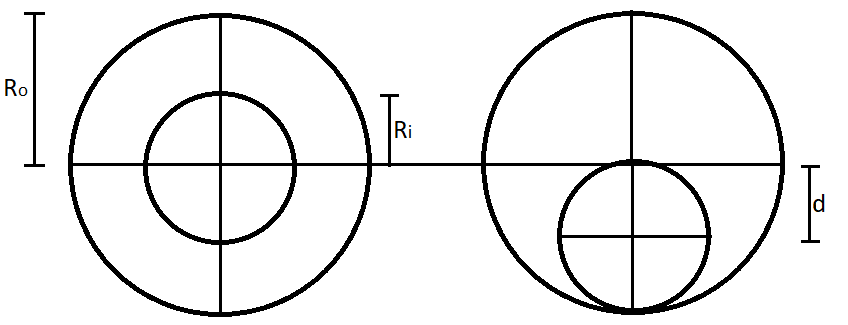}}
\caption{Eccentricity of annulus, $R_{o}$=outer cylinder radius, $R_{i}$=inner cylinder radius, d=distance between cylinder centers}
\end{center}
\label{fig:eccentricity}
\end{figure}
The eccentricity (E) is determined based on the location and dimensions of the two cylinders as described by
 \begin{eqnarray}
  E = \frac{d}{R_{o}-R_{i}}.
  \label{eq:eccentricity}
 \end{eqnarray}
When the annulus is fully concentric, the eccentricity is always $E=0$, while for an entirely eccentric annulus, the ratio is always $E=1.0$.

\begin{figure}[H]
\begin{center}
\begin{tabular}{cc}
200k cells/m  & Cell distortion \\
{\includegraphics[width = 2.5cm, height=5.0cm]{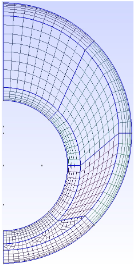}} &{\includegraphics[width = 2.5 cm, height=5.0cm]{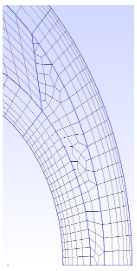}}\\
\end{tabular}
\caption{Cross section (left) and zoomed in view of skewed cells within rotated narrow gap (right)}
\label{fig:mesh}
\end{center}
\end{figure}

We construct the simulation domains in Gmsh. Gmsh is a robust meshing tool which interfaces well with OpenFOAM. Through Gmsh, we define the separate regions of the mesh. The two circles that represent the inner and outer cylinder are subject to the no-slip condition, and they are connected by transfinite lines that originate at predefined points along the circumference. 
Each transfinite line connecting the two cylinders contains three sections. We refer to these sections as the central region and inner and outer wall region.
The wall regions form a significantly refined 5mm thick concentric belt around each cylinder (Fig. \ref{fig:mesh}).

The cylinder wall and wall refinement regions are also transfinite lines. A transfinite line may be used as part of the structural domain, as is the case of the cylinder walls, or solely for its functionality.  The functionality of the transfinite line command is to define the number of mesh points along the line. When properly paired such that each region forms a square of transfinite lines, it gives the user control over the type of element within the region. The drawback of utilizing the transfinite line function is that it prefers having the same number of nodes either side of the line, we consider this to be a soft restriction as it can be bypassed by allowing skewed elements. 

The eccentric configuration results in a narrow gap separating the cylinders, the combination of an eccentric configuration and OpenFOAM's preference for hexahedral elements means we have to choose between having minuscule elements within the gap or circumvent the transfinite line node restriction. We have chosen to reduce the number of elements within the narrow gap by allowing some cell distortion, as shown in figure \ref{fig:mesh}.

Allowing these distorted elements circumvents having the same number of cells in each mesh region within the interior; however, it introduces additional cell skewness and distortion.

Each mesh is quality assured using the built-in commands in OpenFOAM, and they are well within acceptable limits for simulation in terms of non-orthogonality and cell skewness. The maximum and average non-orthogonality for any mesh is 55 and 15 degrees, while the cell skewness maximum is $\sim0.5$. OpenFOAM advice that meshes should be below 70 degrees non-orthogonality and below 4 for cell skewness. Through the case directory, the solver is instructed to perform two specific correction steps to the pressure solution to minimize any effect the non-orthogonality and cell skewness may have on the simulations.

The domain utilized for the horizontal simulations is 7 m long. In the case of the inclined simulation, the domain is 5 m long, and the direction of gravity rotated so that the inclination is 4 degrees. 

Finally, we utilize the mesh in combination with periodic boundary conditions, which allows the flow to transfer seamlessly from the outlet to the inlet. No-slip was applied at the cylinder walls, while we define the centerline as a symmetry plane. We used the k -$\omega$ turbulence model, which is a well-documented closure method for the Reynolds-averaged Navier-Stokes equations.

\section{Fundamental Equations and interFoam} \label{sec:interFoam}
The OpenFOAM solver interFoam is based on a multiphase Volume of Fluid (VOF) model, which utilizes an imaginary mixture fluid instead of solving one equation for each phase. In this work, we introduce a simple modification to the standard interFoam solver to adjust for periodic boundary conditions in an inclined pipe by uncoupling the gravity term from the pressure equation.

A mixture velocity, sometimes referred to as phase-averaged velocity $\bar{u}$, is calculated by applying the following mixture rule

 \begin{eqnarray}
  \bar{u} = (1-\alpha) u_{g}+\alpha u_{l},
  \label{eq:averaging}
 \end{eqnarray}
where $\alpha$ is the phase fraction of liquid in the computational cell. The phase fraction (or indicator function) $\alpha$ is described as

\begin{align}
\alpha=
\begin{cases}
1 & \text{if cell is occupied by liquid}\\
0<\alpha<1 & \text{if cell contains both gas and liquid}\\
0 & \text{if cell is occupied by gas}.
\end{cases}
\label{eq:indicator}
\end{align}
The value assigned to each cell is thus based on the fraction of the fluid contained within this cell. The function returns a value of 1 if the cell contains only liquids and 0 if the cell is filled with gas. The indicator function $\alpha$ is solved for in a modified advection equation

\begin{eqnarray}
 \frac{\partial \alpha}{\partial t}+\nabla\cdot(\alpha\bar{u})+\nabla\cdot(u_{c}\alpha(1-\alpha))=0,
  \label{eq:transport}
\end{eqnarray}
 where the interface compression velocity, $u_{c}$, is used to artificially "compress" the surface, thus maintaining a sharp interface between the two phases. 

With phase-averaging in place, the governing momentum and continuity equations can be written as

\begin{align}
 \frac{\partial \bar{u}}{\partial t} +\nabla\cdot(\bar{u}\bar{u})&=-\frac{1}{\bar{\rho}}\nabla p +  \nabla \cdot (\bar{\nu}(\nabla \bar{u} + (\nabla{\bar{u}})^{T}))+g+\frac{F_{s}}{\bar\rho},
 \label{eq:momentum} \\
 \nabla \cdot \bar{u} &= 0,
\label{eq:continuity}
 \end{align}
where $\bar{\rho}$, $\bar{\nu}$, and $F_{s}$ represent mixture density, viscosity, and surface tension force, respectively. The calculation of the mixture components follows the same mixture rule as exemplified in Eq. \eqref{eq:averaging}.

The benefit of the VOF approach is that the momentum and continuity equations are solved once for the mixture fluid instead of once for each phase. The drawback is that some information about the behavior of each phase is lost. With regards to both the VOF solver and the interface compression, \cite{Desphande} and  \cite{Berberovic} offers an in-depth description.

\section{Experimental setup}
\begin{figure*}[ht]
\begin{center}
{\includegraphics[height = 3.0cm]{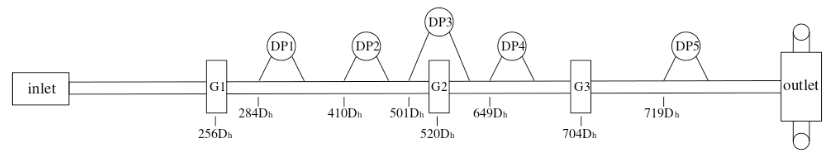}}
\caption{Schematic of Flow loop, Hydraulic diameter (Dh) = 0.05m, G= Gamma densitometer, DP=Differential pressure transducer}
\label{fig:flowloop}
\end{center}
\end{figure*}

We compare the simulations with experimental data gathered from fully concentric and eccentric annuli in a medium scale (45 m long and 99 mm inner diameter) flow loop at IFE. From the experiments, we extract 4 types of data; mixture velocity, instantaneous holdup, pressure gradient, and visual data of the flow field. We define the instantaneous holdup as the volume fraction of liquid within the cross-section. In the case of the experiments, the cross-sectional holdup was determined by using broad beam gamma densitometers (G) at 3 separate locations (Fig. \ref{fig:flowloop}). The gamma densitometers acquire holdup data at 50 Hz and function by measuring the average attenuation of the signal. The attenuated signal is then used to calculate the resulting phase fractions within the sampled cross-section, a simple enough procedure given that we know the total distance travelled and the properties of each fluid.

In general terms, the intensity ($\gamma$) of an incident beam ($\gamma_{o}$), which has passed through a medium is determined by

\begin{eqnarray}
\gamma=\gamma_{o}exp(-\mu t),
\label{eq:intensity}
\end{eqnarray}

where $\mu$ is the attenuation coefficient, and t distance traveled. For two-phase flows, the average cross-sectional holdup is calculated by

\begin{eqnarray}
\alpha_{l}=\frac{log(\frac{\gamma_{m}}{\gamma_{g}})}{log(\frac{\gamma{l}}{\gamma_{g}})}.
\label{eq:HUGD}
\end{eqnarray}

The calculated liquid holdup ($\alpha_{l}$) is thus a ratio of the calibrated single-phase gas and liquid gamma intensities ($\gamma_{g}$,$\gamma_{l}$), as well as the measured gamma intensity ($\gamma_{m}$). The calibrated intensities are determined by single-phase measurements of the gamma beam attenuation, as described by Eq. \eqref{eq:intensity}. As the incident gamma beam passes through a fluid, the radiation intensity of the signal is reduced exponentially as a function of distance traveled through the fluid and the attenuation coefficient. The beam attenuation during the two-phase experiments determines the measured gamma intensity. 

The pressure gradient was measured at 5 separate locations by differential pressure transducers (DP) along the top of the pipe. The DP consist of two pressure measurement devices separated by 3 m with an acquisition rate of 1.2 Hz. By measuring the pressure instantaneously at two locations, the pressure gradient is simply the difference divided by the distance between the measurement devices. The simulations, on the other hand, calculate the required pressure gradient to drive the flow. The calculation is performed as a corrector to the momentum equation and is independent of domain length.

One unfortunate side effect of the low sampling frequency of the experimental pressure gradient is that it does not match the experiment holdup or the simulation data, which both sample at 50 Hz. The low fidelity of the experiment pressure data makes it complicated to compare experiment and simulation with regards to pressure behavior. The obvious solution would be to down-sample the simulation data to match the sampling frequency of the experiments (1.2 Hz). However, for a time-series of 25 s, the simulations would then only have 30 unique data points.

\section{Results}

\begin{table}[H]
\caption{Fluid properties, horizontal annulus.}
\begin{center}
\begin{tabular}{|c|c|c|}
\hline
    Property & Magnitude & unit\ \\
\hline \hline
$\nu_{l}$ & $1.75\cdot10^{-6}$ & $\frac{m^2}{s}$ \\
\hline
$\nu_{g}$ & $6.2\cdot10^{-7}$ & $\frac{m^2}{s}$  \\
\hline
$\rho_{l}$ & 801.0 & $\frac{kg}{m^3}$ \\
\hline
$\rho_{g}$ & 24.30 & $\frac{kg}{m^3}$ \\
\hline
$\sigma$ & 0.0285 & \\
\hline
\end{tabular}
\label{tab:exp_sim_setup}
\end{center}
\end{table}

The fluid properties of the liquid and gas phases are consistent across each horizontal case (Fig. \ref{tab:exp_sim_setup}). However, phase fractions and mixture velocities are altered to induce different flow regimes. The mixture viscosity and density can be determined using Eq. \eqref{eq:averaging}.
Each concentric experiment has an eccentric counterpart of the same mixture velocity but heightened liquid phase fraction.  In order to exemplify the difference that occurs when the holdup fraction is changed, we have simulated all the experiment cases in a partly eccentric domain ($E=0.5$). We mention the phase fractions and mixture velocities for each case in their respective sections. We also summarize the fluid properties for the inclined experiment and simulation in the appropriate section.

 \subsection {Experiment results}
 The experimental results presented in this section fall under 3 separate categories,
 horizontal concentric, horizontal eccentric, and inclined eccentric. 
 The concentric and eccentric horizontal experiments are similar, in that they have the same mixture velocities, but different phase fractions and eccentricities.  The averaged mixture velocity and fractional holdup data gathered in experiments serve as starting points for the simulations. Furthermore, we compare the experimental pressure gradient and instantaneous cross-sectional holdup with the simulations.
 
 \subsubsection{Horizontal concentric annulus experiments}
 As previously mentioned, the horizontal experimental cases are conducted both in a concentric annulus and fully eccentric annulus, with the same mixture velocities. We summarize the mixture velocities and phase fractions for the concentric cases in Tab. \ref{tab:3series:flow:summary} \footnote{Case \# 1,2,3 and 4 correspond to experiment \# 3051, 3054, 3102 and 3117 in the IFE experiment database}. As reported by \cite{Ibarra}, the absolute measurement uncertainty for this particular experimental setup is $\pm$ 1.5 \%.
 Note that the latter two cases employ a significantly higher mixture velocity than the first two cases. 
 
  \begin{table}[H]
\caption{Flow summary, horizontal concentric experiment cases.}
\begin{center}
\begin{tabular}{|c|c|c|}
\hline
Case  \# & $U_{mix}$ (m/s)  & $\alpha$ (\%) \\
\hline\hline
1 & 1.20 & 44.75 \\
\hline
2 & 2.70 & 23.19   \\
\hline
3 & 4.20 & 38.56   \\
\hline
4 & 4.10 & 46.62   \\
\hline
\end{tabular}
\label{tab:3series:flow:summary}
\end{center}
\end{table}

 \begin{figure}[ht]
\begin{center}
{\includegraphics[width = 7.0cm]{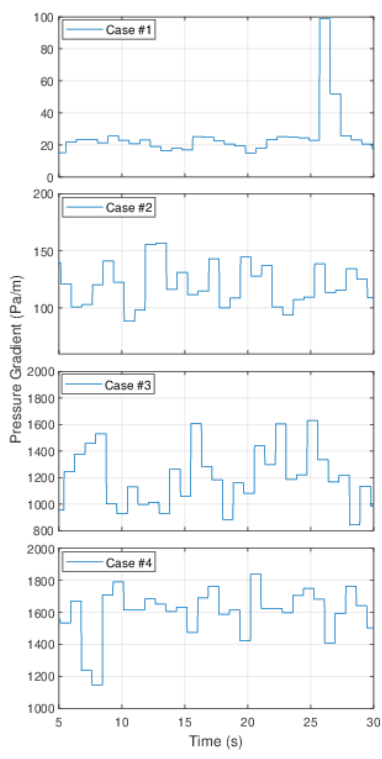}}
\caption{Pressure gradient as a function of time, horizontal concentric  experiment cases}
\label{fig:PVT_horiz_exp_3series}
\end{center}
\end{figure}
 As shown in Fig. \ref{fig:PVT_horiz_exp_3series}, there are two cases with low pressure gradient and two cases with high pressure gradient. The cases with low pressure gradients are intuitively believed to be wavy flow cases (possibly with a single slug in case 1), while the high pressure gradient cases are likely to experience slugs. The highly fluctuating behavior displayed in cases 3 and 4 are indicative of slug flow; however, in order to ascertain the flow regime, the pressure data is supplemented with cross-sectional holdup data. The entire data set for the pressure gradient readings consists of 100 seconds and 120 independent measurements. Tab. \ref{tab:3series:pa:exp} presents a summary of the concentric horizontal annulus pressure gradient data.
 
 \begin{table}[H]
\caption{Pressure gradient summary, horizontal concentric experiment cases.}
\begin{center}
\begin{tabular}{|c|c|c|c|}
\hline
Case \# & 5\% (Pa/m)  & mean (Pa/m) &  95\% (Pa/m)  \\
\hline\hline
1 & 16.80 & 31.56 &  67.92 \\
\hline
2 & 84.59 & 118.22 & 149.28  \\
\hline
3 & 782.62 & 1196.49  & 1624.38  \\
\hline
4 & 1179.12 & 1560.04 & 1789.92  \\
\hline
\end{tabular}
\label{tab:3series:pa:exp}
\end{center}
\end{table}
 
\begin{figure}[ht]
\begin{center}
{\includegraphics[width = 7.0cm]{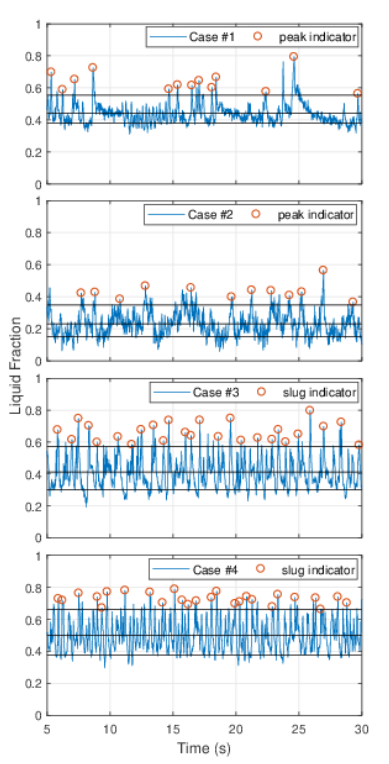}}
\caption{Holdup as a function of time, horizontal concentric experiment cases}
\label{fig:HUVT_3series}
\end{center}
\end{figure}

 In Tab. \ref{tab:3series:pa:exp}, we have separated the data into three columns corresponding to the 5 and 95 \% thresholds as well as the mean value. The 5 and 95 \% thresholds indicate the pressure gradients, where 5 \% of the data is either below or above the respective threshold. These serve as an estimator for the minimum and maximum values of the pressure gradient. We use the thresholds in place of global maximum and minimum values to alleviate the likelihood of including an extreme outlier. In combination with the pressure data, the holdup data (Fig. \ref{fig:HUVT_3series}) is analyzed to determine the flow regime of each case.

We accompany the experiment cases shown in Fig. \ref{fig:HUVT_3series} by three horizontal lines, each of which represents a meaningful metric. The central line is the overall mean holdup, while the upper and lower lines represent the fractional holdup determined as the threshold for a likely slug and likely bubble for the slug cases. The slug indicators are constructed by following the slug identification procedure utilized in \cite{Nuland}. The method builds upon first determining the mean holdup; we then classify the data as above or below the mean. The data above the mean functions to determine a threshold for a slug. The data below the mean is used to determine a bubble threshold, which separates slugs, in order for the indicator to identify a new slug, the profile must have transitioned through the bubble-slug-bubble criterion. The horizontal lines help illustrate the larger intermittent waves for the wavy flow cases.

There are indications that the large waves seen in case \#1 (Fig. \ref{fig:HUVT_3series}) arrive in packets, as seen in the period between 15 and 20 s. Case \#2, on the other hand, displays more evenly distributed occurrences of large waves, and they appear reasonably periodic in time without any significant discrepancies.

 \begin{table}[H]
\caption{Horizontal concentric annulus experiments, slug and wave frequencies.}
\begin{center}
\begin{tabular}{|c|c|c|}
\hline
Case \# & Slug frequency (Hz) & Wave frequency (Hz)  \\
\hline\hline
1 & - & 0.47    \\
\hline
2 & - & 0.52    \\
\hline
3 & 1.06 & -   \\
\hline
4 & 1.09 & -  \\
\hline
\end{tabular}
\label{tab:3series:hz:exp}
\end{center}
\end{table}

The holdup data indicate that the flow regime of the two low velocity cases (1 \&  2) are consistent with high frequency waves. Case 2 behaves more regular in terms of the wave amplitude as the liquid holdup varies between 0.15 and 0.4. Case 1 consists of high frequency small waves, with a few intermittent large waves, while cases 3 and 4 experience slugs, with a slug frequency of 1.06 - 1.09 Hz. Note that the slug and wave frequencies summarized in Tab. \ref{tab:3series:hz:exp} are extracted from the entire 100 s time series, while we show only 25 s for illustration purposes.

\begin{figure}[H]
\begin{center}
\begin{tabular}{cccc}
Case \#1\\
{\includegraphics[width = 8.0cm, height=1.5cm]{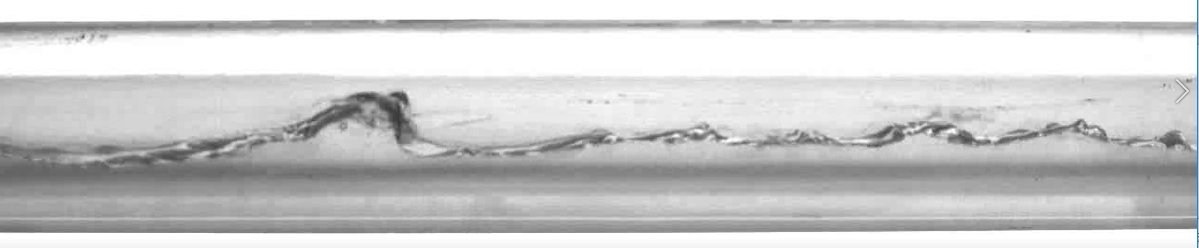}} \\
Case \#2\\
{\includegraphics[width = 8.0 cm, height=1.5cm]{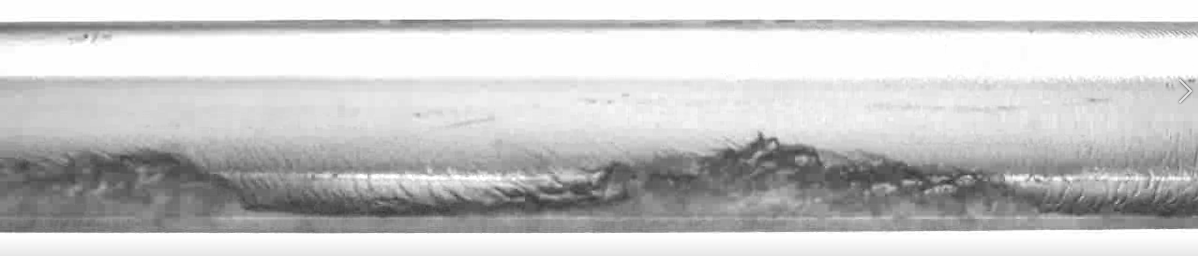}}\\
Case \#3\\
{\includegraphics[width = 8.0 cm, height=1.5cm]{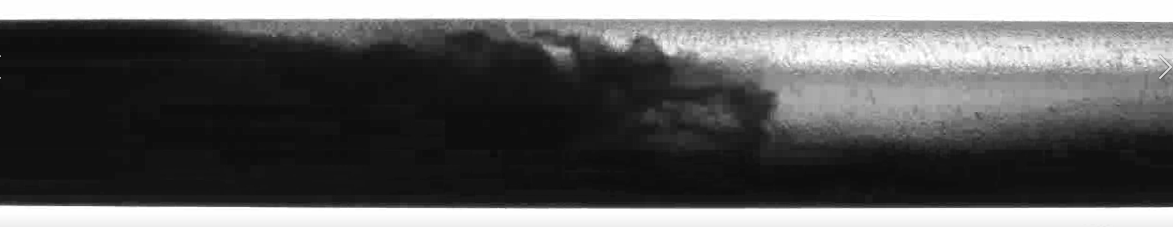}}\\
Case \#4\\
{\includegraphics[width = 8.0 cm, height=1.5cm]{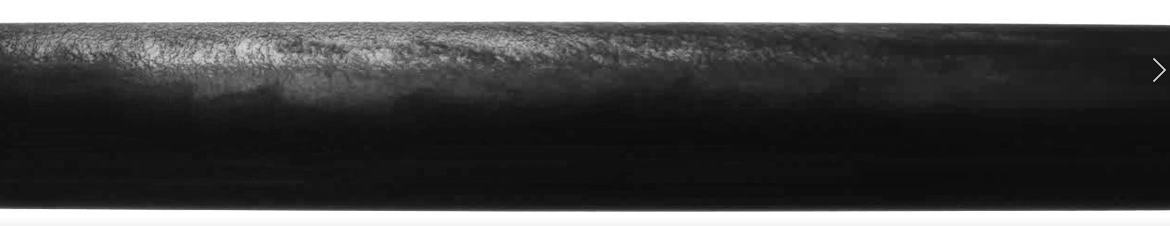}}\\
\end{tabular}
\caption{Snapshots of flow regime for horizontal concentric annulus experiments}
\label{vis:dat:3series}
\end{center}
\end{figure}

 Alongside the fractional holdup data, the experiments also utilized various cameras located along the test section. The location of one of these image-capturing stations was 35 m downstream of the inlet. The visual data at this test section corroborates the flow regimes identified from the holdup data, as shown in Fig. \ref{vis:dat:3series}.

We represent each case with a snapshot of the observed flow regime (Fig. \ref{vis:dat:3series}). The snapshot for experiment case 1 shows several small waves followed by a larger breaking wave. The image for case 2 shows two relatively large waves following each other, while cases 3 and 4 both represent slug flow. 
Although the liquids appear visually different in each snapshot, the fluids present are the same hydrocarbons in each case. Small bubbles of gas permeating through the liquid layer cause the observed darkening effect of the liquid, the effect is worsened in the two slug flow cases.

\subsubsection{Horizontal eccentric annulus experiments with E=1.0}
The important flow properties which we extract from the experimental data and utilize as initialization values for the related simulations (Sim. case 5-8) are summarized in Tab. \ref{tab:6series:flow:summary}.
  \begin{table}[H]
\caption{Flow summary, horizontal eccentric annulus experiment  cases (E=1.0).}
\begin{center}
\begin{tabular}{|c|c|c|c|}
\hline
Case \# & $U_{mix}$ (m/s)  & $\alpha$ \%  \\
\hline\hline
5 & 1.20 & 62.54 \\
\hline
6 & 2.70 & 47.58  \\
\hline
7 & 4.20 & 44.95   \\
\hline
8 & 4.10 & 53.49  \\
\hline
\end{tabular}
\label{tab:6series:flow:summary}
\end{center}
\end{table}

 The horizontal eccentric annulus experiments \footnote{Case \# 5,6,7 and 8 correspond to experiment \# 6005, 6008, 6089 and 6106 in the IFE experiment database.} are similar to their concentric counterparts. The mixture velocities are consistent with the concentric annulus cases. However, they differ in that the liquid phase fractions are higher than their concentric counterparts; also, the inner cylinder is fixed against the bottom of the outer cylinder ($E=1.0$).

\begin{figure}[H]
\begin{center}
{\includegraphics[width = 8.0cm]{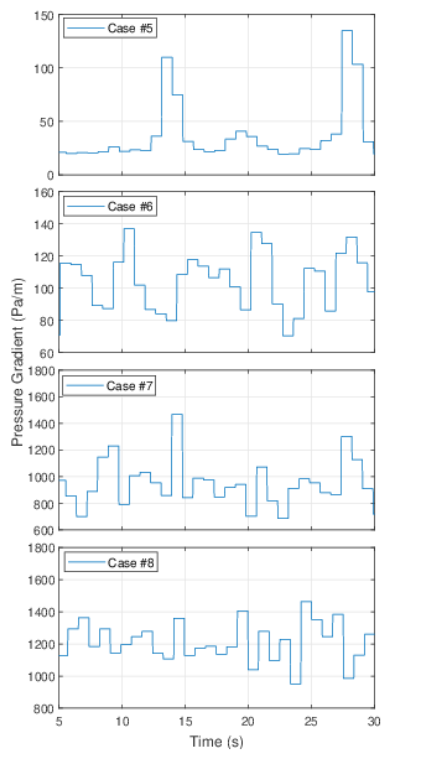}}
\caption{Pressure as a function of time, horizontal eccentric annulus experiment cases (E=1.0).}
\label{fig:PVT_horiz_exp}
\end{center}
\end{figure}
At first glance, the eccentric (Fig. \ref{fig:PVT_horiz_exp_3series}) and concentric (Fig. \ref{fig:PVT_horiz_exp}) pressure gradient measurements appear to be similar. Cases 1 and 5 are both at 1.2 m/s mixture velocity, and display similar pressure gradient trends, with a low overall pressure gradient accompanied by sudden spikes. The pressure gradient spikes are several times higher than the normal behavior and are likely an effect of large waves. 

Case \#6, which corresponds to case \#2, also appears to behave similarly concerning the pressure gradient. There is a slightly higher average pressure gradient in the case of the concentric experiment; however, the amplitudes of the individual pressure spikes are similar.

There are no discernible changes in the measured pressure gradients between the eccentric and concentric experiments that would indicate a significant difference in the flow regime between cases of the same mixture velocity. Cases 5 and 6 appear to behave as wavy flow, while cases 7 and 8 are reminiscent of slug flow, as seen by the highly fluctuating pressure behavior. The slug flow cases display pressure spikes of several hundred Pa/m in the span of a single timestep. These observations follow the trends observed for the corresponding concentric experiments (Fig. \ref{fig:PVT_horiz_exp_3series}). The only noticeable difference is that there appears to have been a pressure gradient decrease between the concentric cases 2, 3, and 4 and the eccentric cases 6, 7, and 8. Tabs. \ref{tab:6series:pa:exp} and \ref{tab:3series:pa:exp} present an overview of the pressure gradient data for the horizontal eccentric annulus experiment cases.

 \begin{table}[H]
\caption{Pressure gradient summary, horizontal eccentric annulus experiment cases (E=1.0).}
\begin{center}
\begin{tabular}{|c|c|c|c|}
\hline
Case \# & 5\% (Pa/m)  & mean (Pa/m) &  95\% (Pa/m)  \\
\hline\hline
5 & 15.29 & 36.27 &  98.20 \\
\hline
6 & 74.90 & 102.95 & 136.95  \\
\hline
7 & 696.43 & 961.00  & 1332.89  \\
\hline
8 & 1031.45 & 1207.22 & 1405.03  \\
\hline
\end{tabular}
\label{tab:6series:pa:exp}
\end{center}
\end{table}
Relative to their concentric counterparts (Fig. \ref{tab:3series:pa:exp}), the eccentric cases 6, 7, and 8 (Fig. \ref{tab:6series:pa:exp}) have undergone a 20\% reduction to the mean pressure gradient. Compared to a concentric annulus, a fully eccentric configuration has a reduced friction factor through a combination of factors, one of which is the narrow gap separating the cylinders. The reduction of the friction factor makes the pressure gradient drop a predictable outcome. Interestingly, case  5 is an exception. As noted, there is a phase fraction difference between the eccentric and concentric experiments. The liquid phase fraction increase offers a possible reason for why case  5 is the only case without a reduced mean pressure gradient as there is significantly more liquid in case 5 compared to case 1. In subsequent simulations, we will notice that for every case, an increase in holdup fraction equates to an increased mean pressure gradient.

The threshold values for the respective cases also undergo similar reductions. The consistency of pressure gradient behavior is an indicator but does not confirm that the concentric and eccentric cases are consistent with regards to the flow regime. For confirmation, we have to delve into the holdup data and account for any potential differences.

\begin{figure}[H]
\begin{center}
{\includegraphics[width = 8.0cm]{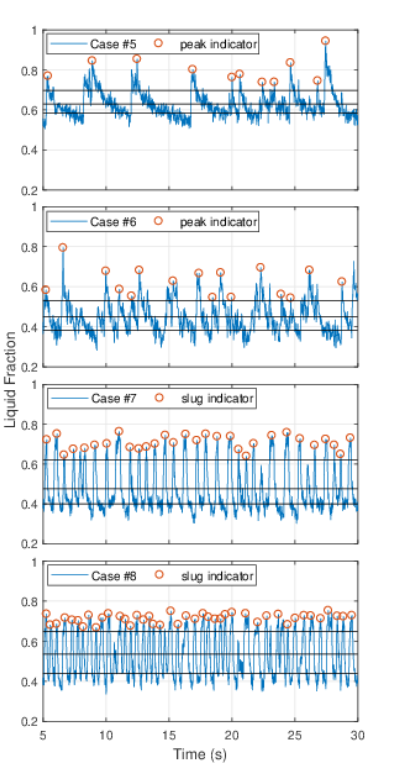}}
\caption{Holdup as a function of time, horizontal eccentric annulus experiment cases (E=1.0)}
\label{fig:hu_horiz_exp}
\end{center}
\end{figure}

As shown in Fig. \ref{fig:hu_horiz_exp}, cases 5 and 6 are fundamentally different to 7 and 8. The two prior cases indicate wavy flow with large intermittent waves. These observations are consistent with the low pressure variations shown in Fig. \ref{fig:PVT_horiz_exp}. For the low velocity flow cases, the increase in average liquid holdup between the concentric and eccentric cases combined with the changed eccentricity results in larger waves. Especially case \#5 (Fig. \ref{fig:hu_horiz_exp}) has less extended periods of ripple waves and more large waves when compared to case \#1 (Fig. \ref{fig:HUVT_3series}). 

 \begin{table}[H]
\caption{Wave and slug frequencies, horizontal eccentric experiment cases (E=1.0).}
\begin{center}
\begin{tabular}{|c|c|c|}
\hline
Case \# & Slug frequency (Hz) & Wave frequency (Hz)  \\
\hline\hline
5 & - & 0.44    \\
\hline
6 & - & 0.64    \\
\hline
7 & 1.16 & -   \\
\hline
8 & 1.64 & -  \\
\hline
\end{tabular}
\label{tab:6series:hz:exp}
\end{center}
\end{table}

Cases 2 and 6 appear similar with no significant change in the holdup pattern, although the large waves are of a higher amplitude. The two slug cases remain sluggish but at an increased frequency (Tab. \ref{tab:6series:hz:exp} \& \ref{tab:3series:hz:exp}). For the experiments, changing the eccentricity and phase fractions while maintaining the same mixture velocity, appear to have had little impact on the resulting flow regime, apart from increasing the slug frequency and wave amplitude.

\begin{figure}[H]
\begin{center}
\begin{tabular}{cccc}
Case \#5\\
{\includegraphics[width = 8.0cm, height=1.5cm]{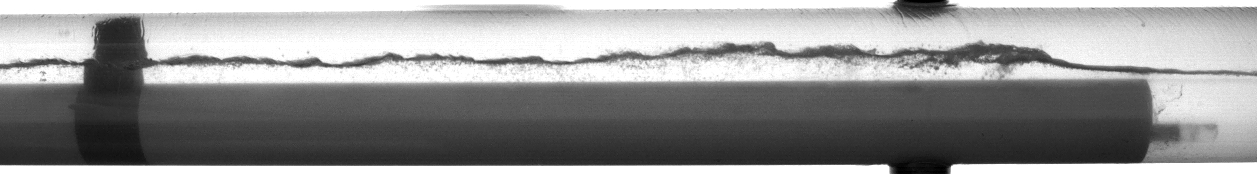}} \\
Case \#6\\
{\includegraphics[width = 8.0 cm, height=1.5cm]{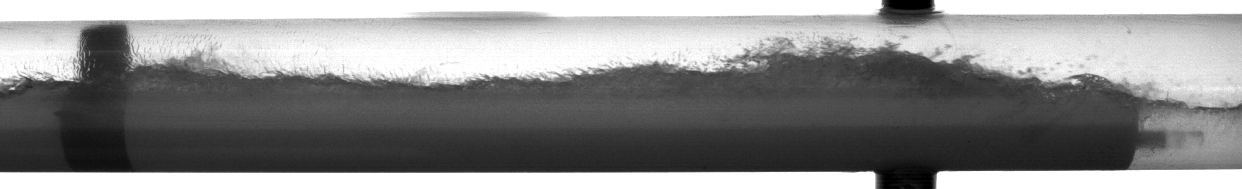}}\\
Case \#7\\
{\includegraphics[width = 8.0 cm, height=1.5cm]{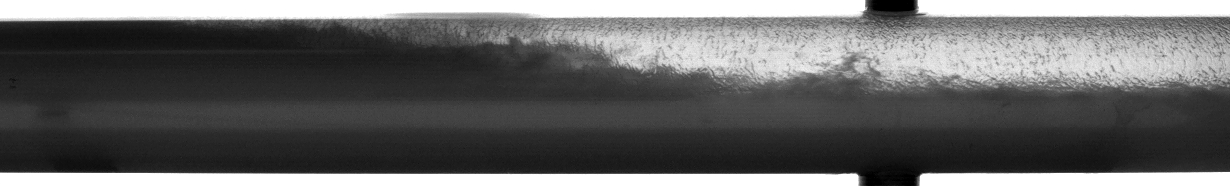}}\\
Case \#8\\
{\includegraphics[width = 8.0 cm, height=1.5cm]{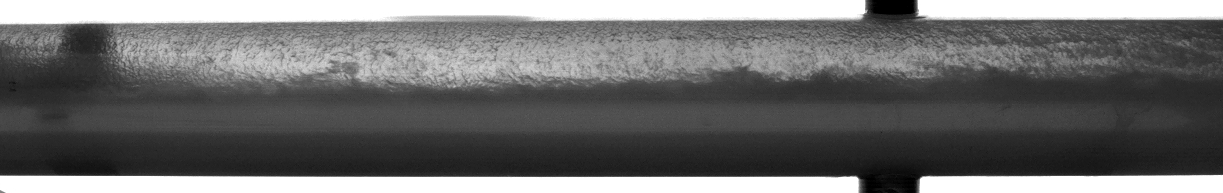}}\\
\end{tabular}
\caption{Snapshots of flow regime for horizontal eccentric annulus with E=1.0}
\label{vis:dat:6series}
\end{center}
\end{figure}
The typically observed flow regimes (Fig. \ref{vis:dat:6series}) of the entirely eccentric experimental cases are similar to the concentric experimental cases (Fig. \ref{vis:dat:3series}). Although each case has a higher average liquid holdup, the flow regimes are consistent. Case \# 5 is perhaps the one exception. The average liquid level means the interface between liquid and gas coincides with the top of the inner pipe.  The result is that there are shorter periods with ripple waves interrupted by large intermittent waves.  Both of these wave types are discernible from the holdup data (Fig. \ref{fig:hu_horiz_exp}); the high amplitude waves are marked by indicators, while the smaller waves are the erratic region following the large waves. 

The two slug flow cases visually behave the same as the concentric cases; however, with an increased slug frequency. While both the concentric cases had a slug frequency of 1.06-1.09 Hz, cases 7 and 8 have slug frequencies of 1.16 and 1.64 Hz, respectively. The observation is interesting and could be an effect of either the increased average holdup or comparing two cases with different eccentricity. Because the simulations all share the same eccentricity, we can use the simulations to ascertain if the eccentricity is contributing to the difference in slug frequency or if it is an effect of the increased holdup between cases 1-4 and 5-8.

\subsubsection{4 degree inclined eccentric annulus experiment}
The $4^\circ$ inclined experiments \footnote{Case \# 9 correspond to experiment \# 7049 in the IFE experiment database} were performed in a completely eccentric configuration (E=1.0), in the same flow loop as the previously presented experiments (Fig. \ref{fig:flowloop}). 
The fluid properties are slightly different than for the horizontal configuration (Fig. \ref{tab:exp_sim_setup}) as summarized in Fig. \ref{tab:exp_sim_setup_inc}.

\begin{table}[H]
\caption{Fluid properties, inclined eccentric experiment with E=1.0}
\begin{center}
\begin{tabular}{|c|c|c|}
\hline
    Property & Magnitude & unit\ \\
\hline \hline
$\nu_{l}$ & $2.6\cdot10^{-5}$ & $\frac{m^2}{s}$ \\
\hline
$\nu_{g}$ & $3.42\cdot10^{-7}$ & $\frac{m^2}{s}$  \\
\hline
$\rho_{l}$ & 854.60 & $\frac{kg}{m^3}$ \\
\hline
$\rho_{g}$ & 43.83 & $\frac{kg}{m^3}$ \\
\hline
$\sigma$ & 0.0285 & \\
\hline
\end{tabular}
\label{tab:exp_sim_setup_inc}
\end{center}
\end{table}

The primary differences between the inclined flow experiments and the horizontal ones are that the liquid phase utilized in the inclined experiments is significantly more viscous, and that the inner pipe had a smaller diameter of 40 mm compared to 50 mm for the horizontal cases. Both fluids used in the inclined experiment are also denser than their horizontal experiment counterparts. The reason for the higher gas density is a higher system pressure in these experiments.

\begin{figure}[H]
\begin{center}
{\includegraphics[width = 8.0cm]{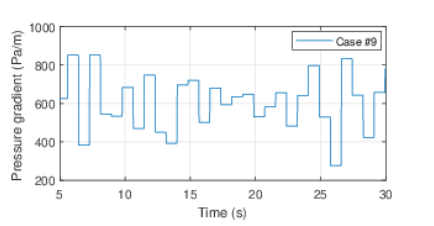}}
\caption{Pressure as a function of time, inclined eccentric experiment with E=1.0}
\label{fig:PVT_horiz_exp_7049}
\end{center}
\end{figure}
The inclined experiment time series lasts for 120 s; however, we show 25 s for illustration purposes (Fig. \ref{fig:PVT_horiz_exp_7049}), while we analyze the entire 120 s for the statistical behavior of the flow. The pressure gradient behavior is reminiscent of large waves or potentially slug flow. Typically, a slug flow at this mixture velocity will have a higher average pressure gradient; however, a low frequency slug flow could potentially behave as shown. The relative difference from the mean to maximum and minimum thresholds are roughly 35 - 45 \% as summarized in Tab. \ref{tab:exp_7049:pa}.

 \begin{table}[H]
\caption{Pressure gradient summary, 4 degree inclined eccentric experiment (E=1.0).}
\begin{center}
\begin{tabular}{|c|c|c|c|}
\hline
Case \# & 5\% (Pa/m) & Mean (Pa/m) & 95\% (Pa/m) \\
\hline\hline
9 & 385.11 & 601.66 & 877.98  \\
\hline
\end{tabular}
\label{tab:exp_7049:pa}
\end{center}
\end{table}

\begin{figure}[H]
\begin{center}
{\includegraphics[width = 8.0cm]{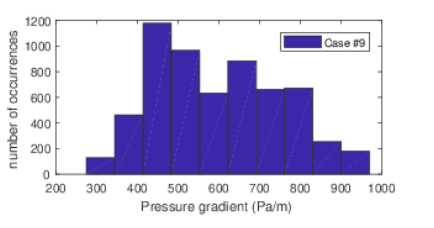}}
\caption{Pressure gradient histogram, inclined eccentric experiment (E=1.0)}
\label{fig:Pa_hist_exp_7049}
\end{center}
\end{figure}

The histogram presents a visual representation of the pressure gradient behavior. As shown in Fig. \ref{fig:Pa_hist_exp_7049}, there are rare occurrences of pressure gradient spikes and troughs at near 1000 and 300 Pa/m. The shape is reminiscent of a Gaussian distribution. However, the left tail abruptly stops while the right tail extends; this is known as a right-skewed distribution.

\begin{figure}[H]
\begin{center}
{\includegraphics[width = 8.0cm]{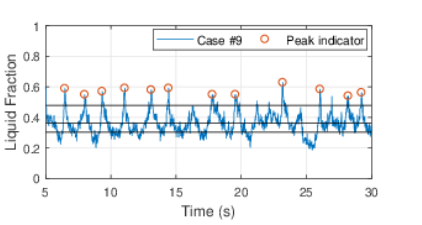}}
\caption{Holdup as a function of time, inclined eccentric experiment with E=1.0}
\label{fig:HU_exp_7049}
\end{center}
\end{figure}

\begin{figure}[H]
\begin{center}
\begin{tabular}{cccc}
Large intermittent wave\\
{\includegraphics[width = 8.0cm]{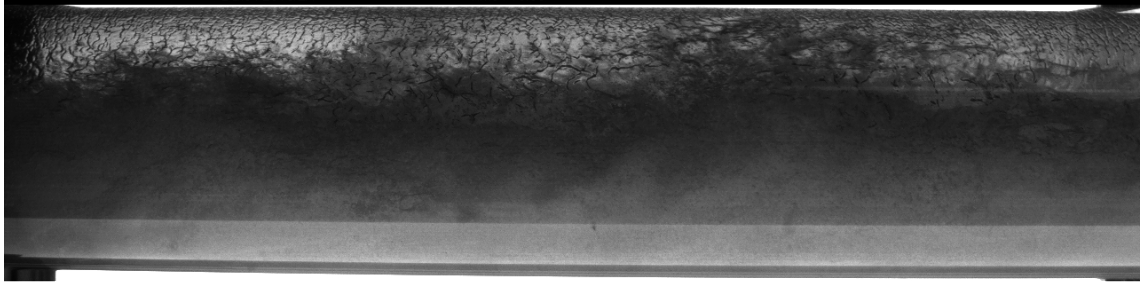}} \\
small frequent waves\\
{\includegraphics[width = 8.0 cm]{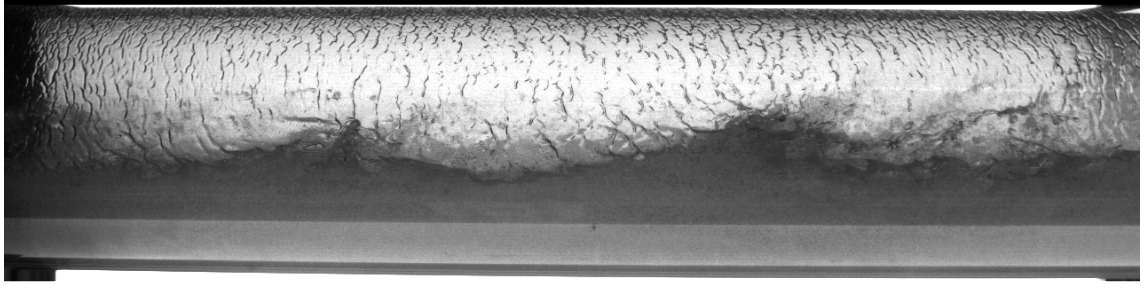}}\\
\end{tabular}
\caption{Snapshots of flow regime for $4^\circ$ inclined eccentric experiment case \# 9 with E=1.0}
\label{vis:dat:inc_exp}
\end{center}
\end{figure}

 The peak holdup readings seen in Fig. \ref{fig:HU_exp_7049} indicate that this is either a significantly aerated slug flow or more likely intermittent large waves. In order to verify the flow regime, we cross-reference with the visual data (Fig. \ref{vis:dat:inc_exp}) captured at the second camera location (Fig. \ref{fig:flowloop}).

The visual data suggests that the peaks indicated in Fig. \ref{fig:HU_exp_7049} are thoroughly aerated liquid waves. Between large liquid waves, there is a period of smaller high frequency waves.  

Note that the band seen at the bottom of each image is not a gap between the inner and outer cylinder but rather an optical disturbance, in fact, the band is visible near the middle of the images for the horizontal cases (Figs. \ref{vis:dat:3series} \& \ref{vis:dat:6series}). There are several problems when taking images of a see-through pipe filled with two fluids. The most likely cause for the bright band is that refraction has caused an optical distortion, which makes it appear that the inner pipe is not touching the outer pipe.

\begin{figure}[H]
\begin{center}
{\includegraphics[width = 8.0cm]{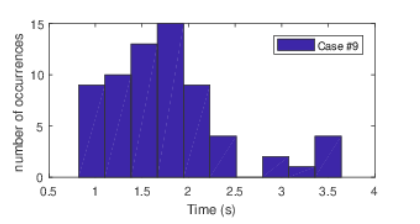}}
\caption{Large wave separation in time, inclined eccentric experiment with E=1.0}
\label{fig:Hz_Exp_7049}
\end{center}
\end{figure}
If we analyze the entire time-series of which 25 s is shown in Fig. \ref{fig:HU_exp_7049}  and determine the period between each peak, we can say something about the wave frequency and frequency distribution. In this case, we have converted the data into a histogram (Fig. \ref{fig:Hz_Exp_7049}).  We notice that the majority of the periods between large waves are in the region from 1 to 2 seconds, the average frequency of large waves is 0.5667 Hz or a wave period of 1.76 s.

\subsection{Simulation results}
Following the previously established convention, we separate the simulation results into three categories, corresponding to the experiment sections. The categories are horizontal eccentric cases 1-4, horizontal eccentric cases 5-8, and inclined eccentric. The horizontal cases are split into two sets corresponding to the experiments they are based on. The first 4 are based off of concentric experiments, while cases 5-8 are based off of fully eccentric experiments. Within each category, we sort the data by holdup and pressure gradient results. As was the case in the experiment section, we will start with cases 1-4. These simulations use the phase fractions and average mixture velocities extracted from the experimental data. However, as mentioned, the simulations are run in a domain with an eccentricity of 0.5. The eccentricity difference allows us to analyze whether it has a significant impact on the flow regime and pressure gradient behavior. 

Based on previous work \citep{Friedemann}, the 7 m domain is the chosen domain length for the simulations.
Although the work showed that shorter domains could be reasonable approximations of experiments, an extended domain minimizes the risk of inadvertently altering the flow by restricting the amount of available liquid within the domain. This effect is more pronounced in strong periodic flows such as slugs or plugs, while wavy flow should be less prone to flow regime deviations caused by domain deficiencies. 

Although the 7 m domain was overall the best representation of both slug frequencies and pressure gradient based on previous simulations, there is an additional computational cost related to a longer domain. To compensate for the prohibitive computational cost, we run the simulations at a lower mesh density than ideally desired. Keep in mind that the previous work indicated that the flow regime is not affected by the mesh density in the range of meshes studied. The pressure gradient is affected to some degree; however, they were shown to be reasonable approximations of the expected result in concentric horizontal configurations. The common trend was that the pressure gradients would be overestimated at low mesh resolution and underestimated at high mesh counts, while flow regime and slug frequencies remained consistent throughout meshes. 

\subsubsection{Horizontal simulations cases 1-4 with eccentricity of 0.5 }
\begin{figure}[H]
\begin{center}
{\includegraphics[width = 8.0cm]{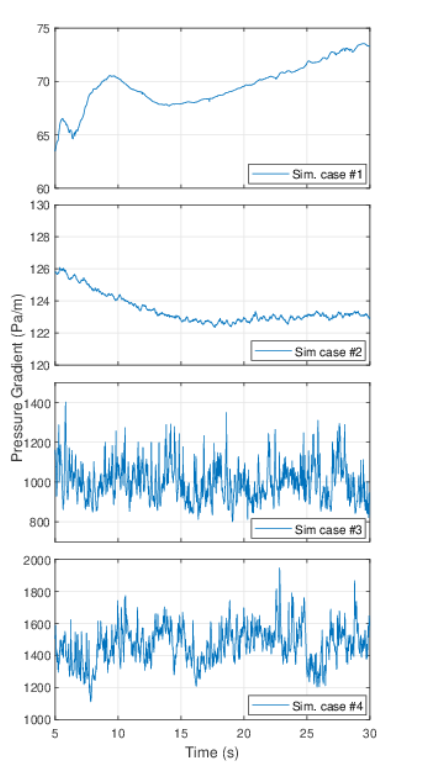}}
\caption{Pressure gradient as a function of time, horizontal eccentric annulus with  E=0.5 and 100k cells/m}
\label{fig:PVT_horiz_ec}
\end{center}
\end{figure}

Figure \ref{fig:PVT_horiz_ec} displays pressure as a function of time for the 4 horizontal cases.  The wavy flow cases (1 \& 2) display a nearly constant pressure gradient state compared to the experiments.  We believe this is a side effect of the periodic boundary conditions combined with the low velocity flow.  The two slug cases (3 \& 4) behave as expected; the periodic build-up and dissipation of slugs result in a violent pressure reaction, varying with several hundred Pa/m  within seconds.

\begin{figure}[H]
\begin{center}
{\includegraphics[width = 8.0cm]{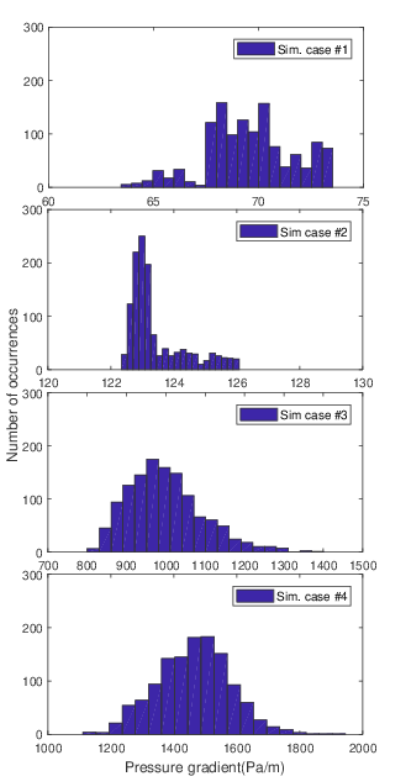}}
\caption{Pressure gradient histogram distribution, horizontal eccentric annulus with  E=0.5 and 100k cells/m}
\label{fig:PVT_horiz_hist_ec}
\end{center}
\end{figure}

The histograms (Fig. \ref{fig:PVT_horiz_hist_ec}) represent a statistical overview of the pressure gradient behavior for simulation cases 1-4.  They show that the slug pressure drop distributions are relatively concentrated. The wavy cases are slightly different, and there are indications that case \# 1 is perhaps not fully developed yet, as the pressure drop seems to be increasing until it drops at the very end. This observation is at odds with previous simulations, which have indicated that after 5 s the flow is developed and in a periodic state.

\begin{table}[H]
\caption{Pressure gradient summary, horizontal eccentric annulus with  E=0.5 and 100k cells/m}
\begin{center}
\begin{tabular}{|c|c|c|c|}
\hline
Sim \# & 5\% (Pa/m)  & mean (Pa/m) &  95\% (Pa/m)  \\
\hline\hline
1   & 65.69  & 69.61 & 73.08 \\
\hline
\hline
2 &  122.60 & 123.45 & 125.51 \\
\hline
\hline
3 &  863.89 & 998.68 & 1168.83  \\
\hline
\hline
4 &  1270.03 & 1464.04 & 1648.63  \\
\hline
\end{tabular}
\label{tab:pressure_info_ec}
\end{center}
\end{table}
Tab. \ref{tab:pressure_info_ec} summarizes the pressure gradient information from simulation cases 1 to 4. We notice that the mean pressure gradients from the simulations are a surprisingly good match to the experimental data, to within 10 \% for the three latter cases. We emphasize that these simulations are not supposed to be an exact match to the experiments as the inner cylinder location does not match the experiment location. Instead, they are meant to highlight changes to the flow behavior that may occur due to the eccentricity difference. 

We notice that for case 1, there is a quite significant mean pressure gradient change. For this particular case, the experiment has an unbroken liquid surface below the concentric pipe when there are no waves, while for the eccentric simulation, the interface between fluid and gas occurs in the presence of the interior pipe. The different eccentricities mean the interfacial area between liquid and gas is more significant in the experiments, while the interface between the cylinder wall and liquid is higher in the simulations, which in turn alters the frictional forces as well as the transfer of energy between phases. 

We have previously worked on similar cases at similar mesh densities and established that the simulations tend to underestimate the maximum pressure gradient by roughly 10 \% and overestimate the minimum and mean pressure gradients by 23.7 \% and 2.3 \% respectively at 100k cells/m. If we similarly determine the relative error compared to the experiments, we see a major difference between the concentric experiments and partly eccentric simulations for case 1, as shown in Tab. \ref{tab:rel_error_100k_case1_4}.

\begin{table}[H]
\caption{Relative error of simulations with respect to concentric experiments for horizontal simulations at E=0.5 and 100k cells/m}
\begin{center}
\begin{tabular}{|c|c|c|c|}
\hline
Sim \# & 5\% (\%)  & mean (\%) &  95\% (\%)  \\
\hline\hline
1   & 291.0  & 120.5 & 7.6 \\
\hline
\hline
2 &  44.93 & 4.4 & -15.92 \\
\hline
\hline
3 &  10.4 & -16.53 & -28.05  \\
\hline
\hline
4 &  7.71 & -6.2 & -7.89  \\
\hline
\end{tabular}
\label{tab:rel_error_100k_case1_4}
\end{center}
\end{table}

\begin{figure}[H]
\begin{center}
{\includegraphics[width = 8.0cm]{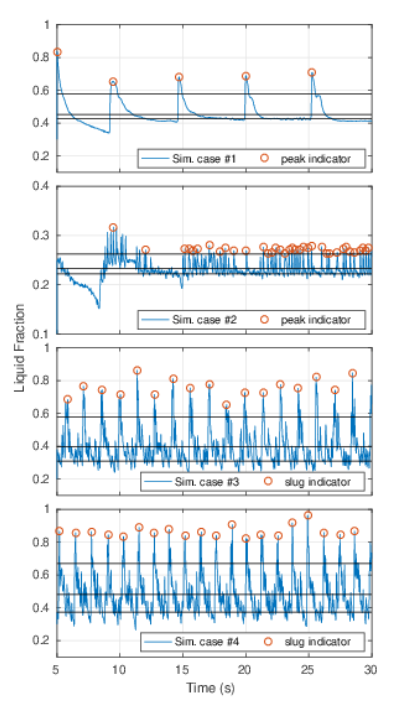}}
\caption{Holdup as a function of time, horizontal eccentric annulus with E=0.5 and 100k cells/m}
\label{fig:hu_horiz_ec_hu_exp}
\end{center}
\end{figure}

The deviations compared to the experiments are as high as 291 \% for case 1.  These behavioral changes could occur due to the narrow gap below the inner cylinder. Alternatively, the cause is the altered location of the interface between fluids and the inner pipe. Considering that the largest differences occur in the wavy flow cases, it is also possible that the computation of the pressure gradient for wavy flow is extremely dependent on both holdup rates and wave type.  As shown in Figs. \ref{fig:hu_horiz_ec_hu_exp} and \ref{fig:HUVT_3series} the wave behavior is significantly different in the simulated case with E=0.5 compared to the experiment with E=0.0. We believe the different flow behavior is responsible for the majority of the pressure gradient shift.
There is also some small error involved in the measurement of mean holdup in the experiments, the discrepancy between actual experiment holdup and simulation holdup could contribute to the computed relative difference.

The remaining cases are in better agreement, yet the errors are still as high as  45 \% for case 2, and below 30 \% for the two slug cases. We notice that for the slug cases, the general trend is that the difference between simulation and experiment is shifted negatively compared to the previous work. The negative shift coincides with the idea that the pressure gradient should decrease in a more eccentric configuration.

\begin{figure}[H]
\begin{center}
{\includegraphics[width = 8.0cm]{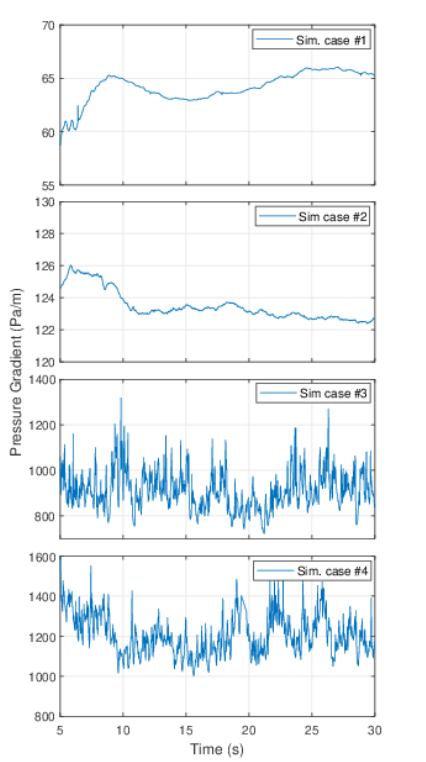}}
\caption{Pressure gradient as a function of time, horizontal eccentric annulus with  E=0.5 and 200k cells/m}
\label{fig:PVT_horiz_me}
\end{center}
\end{figure}

As we refine the mesh (Fig. \ref{fig:PVT_horiz_me}), we notice that there are some slight differences to the pressure gradient behavior (Tab. \ref{tab:pressure_info_me}).

Both the low velocity cases (1 \& 2) agree with the behavior seen in the coarser mesh in terms of the relatively constant pressure gradient, although slightly decreased for simulation case 1. Also, there is a slight change in the overall behavior of the first case. Instead of the weakly increasing pressure gradient shown for the coarser mesh (Fig. \ref{fig:PVT_horiz_ec}, case 1), the pressure gradient flattens out at 65 Pa/m after a period of slowly increasing and decreasing over time. 

\begin{figure}[H]
\begin{center}
{\includegraphics[width = 8.0cm]{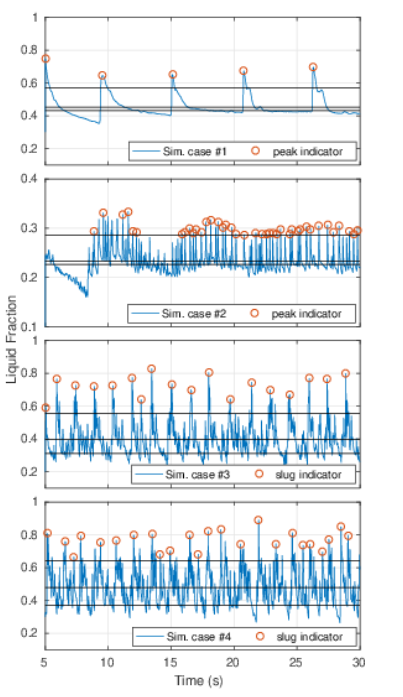}}
\caption{Holdup as a function of time, horizontal eccentric annulus with E=0.5 and 200k cells/m}
\label{fig:hu_horiz_me_hu_exp}
\end{center}
\end{figure}

Case 2 displays the same type of behavior in both meshes, while the two slug cases are at a slightly lower overall pressure gradient. The trend that the pressure gradient decreases with increased mesh density was also noted in \cite{Friedemann} for slug cases. We believe that this occurs due to an inability to resolve minor bubbles and finer flow structures. It is well known that an inability to resolve the minor structures will alter the turbulent field through the energy cascade. Thereby also how turbulence is treated at the fluid interface and wall regions. The effective viscosity is also affected through a lack of mixing. Additionally, the difference in wave amplitude observed for the two case \#2 meshes (Figs. \ref{fig:hu_horiz_ec_hu_exp} \& \ref{fig:hu_horiz_me_hu_exp}) likely contributes to why the finer mesh does not show a reduced mean pressure gradient.

\begin{figure}[H]
\begin{center}
{\includegraphics[width = 8.0cm]{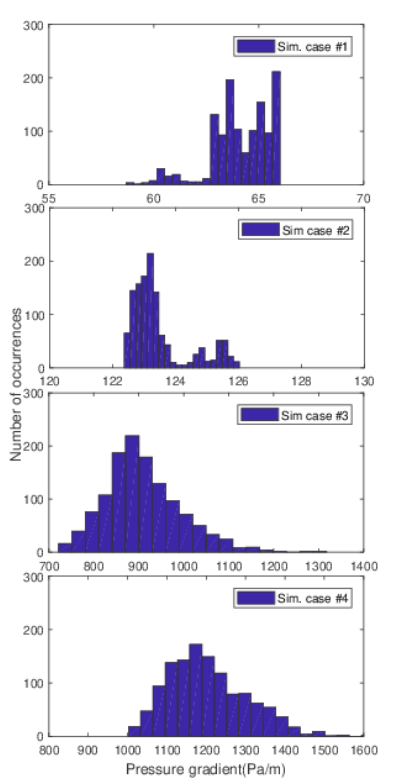}}
\caption{Pressure gradient summary, horizontal eccentric annulus with  E=0.5 and 200k cells/m}
\label{fig:PVT_horiz_hist_me}
\end{center}
\end{figure}

\begin{table}[H]
\caption{200k cells/m mesh pressure summary.}
\begin{center}
\begin{tabular}{|c|c|c|c|}
\hline
Sim \# &  5\% (Pa/m) &  mean (Pa/m) & 95 \% (Pa/m) \ \\
\hline\hline
1 & 60.91 & 64.15 & 65.87   \\
\hline

2 & 122.54 & 123.47 & 125.53  \\
\hline
3 & 785.07 & 909.25  & 1064.34  \\
\hline
4 & 1063.04 & 1203.92 & 1383.30   \\
\hline
\end{tabular}
\label{tab:pressure_info_me}
\end{center}
\end{table}

Comparing the histograms, Figs. \ref{fig:PVT_horiz_ec} \& \ref{fig:PVT_horiz_me}, and Tabs. \ref{tab:pressure_info_ec} \& \ref{tab:pressure_info_me}, for the two meshes, there is one noticeable trend; as we increase the mesh density, the computed pressure gradients decrease. While the minimum values are near the experiment results, the mean and maximum pressure gradient estimator is considerably off for the slug flow cases. The reason the minimum results are reasonable is likely a combination of effects and hard to identify. There are several probable contributors; the overall level of turbulence in the calmer state, the transference of turbulent energy at the interface, and the dampening at the wall. In combination with these effects, during the calmer flow state, there are fewer bubbles present in the experimental measurements, which is more representative of the behavior of the simulations. Conversely, the inability to resolve minor bubbles results in large bubbles forming near the top of the slugs, reducing the required pressure gradient to drive the flow, which affects the peak values of the simulations.  

Previous work \citep{Friedemann} indicates an error estimate of roughly 21 and 28\% undershoot of the mean and maximum pressure gradient at this mesh density, as well as an expected  overshoot of the minimum pressure gradient of 3.4 \%  at 200k cells/m. 

\begin{table}[H]
\caption{Relative difference of simulations with respect to concentric experiments for horizontal simulations at E=0.5 and 200k cells/m.}
\begin{center}
\begin{tabular}{|c|c|c|c|}
\hline
Sim \# &  5\% (\%) &  mean (\%) & 95 \% (\%) \ \\
\hline\hline
1 & 262.6 & 103.3 & -3.0   \\
\hline
2 & 44.9 & 4.4 & -15.9  \\
\hline
3 & 0.3 & -24.0  & -34.5  \\
\hline
4 & -9.8 & -22.8 & -22.7   \\
\hline
\end{tabular}
\label{tab:rel_error_200k_case1_4}
\end{center}
\end{table}

Comparing the corresponding relative differences for the 100  and 200k cells/m meshes (Tabs.\ref{tab:rel_error_100k_case1_4} \& \ref{tab:rel_error_200k_case1_4}), we notice a persistent negative shift for all cases apart from Sim. 2. We attribute the discrepancy to the altered flow pattern of that particular case, as shown in Figs.\ref{fig:hu_horiz_ec_hu_exp} \& \ref{fig:hu_horiz_me_hu_exp}. The presence of larger waves coincides with an increase of turbulent kinetic energy and, thereby, an increased pressure gradient.    The two slug cases behave relatively similar to the experiments; however, consistently undershoots the experiment pressure gradient. The behavior is expected as the friction factor is lower in any eccentric configuration compared to a concentric configuration.

Concerning the holdup profile for the coarsest mesh (Fig. \ref{fig:hu_horiz_ec_hu_exp}) compared to the experiment data (Fig.\ref{fig:HUVT_3series}), there is one significant difference. Namely, case 1 has a complete lack of minor waves or ripple waves.  We believe numerous aspects could cause the effect. One of which is a poor choice of initial conditions, which could induce a near steady-state flow regime that is not the expected physical result. Conversely, the low velocity cases, in particular case 1, takes a significant time to develop; by this logic, the simulation could yet be developing. Another possible cause is that the different eccentricity results in large waves with less small wave effects. Interestingly, there are indications that the wavy flow eccentric experiment cases (Fig. \ref{fig:hu_horiz_exp}) are showing this type of behavior, as shown by the more prominent large waves with the gradual drop off of holdup in between large waves. 

The other cases are more consistent with their experimental counterparts.  Case 2 is exhibiting more uniform waves than that seen in the experiments and also at a significantly higher frequency after 15 s (2.2 Hz). The increased frequency is thought to be caused by the intentional difference between simulation and experimental setup with regards to the eccentricity for these cases. The slug flow cases 3 and 4 are also represented as proto-slug flow by the simulations at a reduced slug frequency of 0.68 and 0.8 Hz compared to 1.06 and 1.09 Hz. 

Compared to the concentric experiments, we have noticed both a change in the wave pattern for cases 1 and 2 as well as slug frequency (cases 3 \& 4). Accompanying these holdup pattern changes, we observe, in some cases, drastic pressure gradient changes, which to us indicates that altering the eccentricity from E=0 to E=0.5 has had an immediate effect on the resultant flow.

The simulations for the increased mesh density (Fig. \ref{fig:hu_horiz_me_hu_exp}) show one noticeable difference compared to the lower density simulations (Fig. \ref{fig:hu_horiz_ec_hu_exp}). The waves produced in the second simulation case are significantly larger. The wave structures have an amplitude of \~ 5 \% liquid holdup, while for the coarser mesh, they are nearer to \~2.5 \%. 

Concerning the comparison between experiments and simulations, we again observe that simulation case 2 maintains an increased wave frequency. If we exclude the initial 15 s where there appears to be some transient behavior, the wave frequency is 2.47 Hz. The frequency is similar to the 100k cells/m simulation (2.2 Hz), but a significantly increased deviation compared to the experiment (0.64 Hz). Both mesh densities indicate that the eccentricity of the annulus has had a significant impact on the flow regime of this case. While not transitioning to a different flow regime, the wave behavior, amplitude, and frequency have changed. 

For the 3 remaining cases, the finer mesh remains consistent with the 100k cells/m simulation with regards to the holdup pattern. The wave frequency and amplitude remain similar (case 1) while we note a small increase in the slug frequency from 0.68 and 0.78 Hz to 0.72 and 0.96 for case 3 and 4, respectively.

\begin{figure}[H]
\begin{center}
\begin{tabular}{cccc}
Sim. case \#1\\
{\includegraphics[width = 8.1cm]{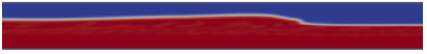}} \\
Sim. case \#2\\
{\includegraphics[width = 8.0 cm]{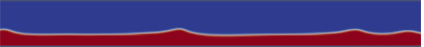}}\\
Sim. case \#3\\
{\includegraphics[width = 8.0 cm]{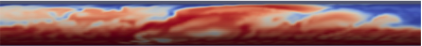}}\\
Sim. case \#4\\
{\includegraphics[width = 8.1 cm]{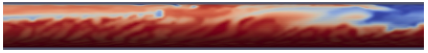}}\\
\end{tabular}
\caption{Snapshots of flow regime, horizontal eccentric annulus with E=0.5 and 200k cells/m}
\label{vis:dat:3series:sim}
\end{center}
\end{figure}
When we compare the snapshots of the eccentric simulation cases 1 to 4 (Fig. \ref{vis:dat:3series:sim}) with their corresponding experiment images (Fig. \ref{vis:dat:3series}), the most immediate observation is the altered flow state of simulation case 1 compared to the experiment. We do not observe the long rolling waves observed in the experiments. The long waves observed in the experiments are usually both preceded and followed by a region of smaller waves reminiscent of ripples. The reason there are no ripple waves in the simulation is likely two-fold. A combination of the initial conditions and the mesh density can produce unexpected flow patterns. In the case of ripple waves, if the mesh is not able to resolve them, the complete removal is a possible outcome.  For the remaining cases, the observations are more subtle and better observed by the holdup fractions.

\subsubsection{Horizontal simulation cases 5-8}

Simulation cases 5 to 8 were only run in the 100k cells/m mesh, as the flow regime shows a limited change between meshes for cases 1 to 4. Additionally, the pressure results are of comparable values.

\begin{figure}[H]
\begin{center}
{\includegraphics[width = 8.0cm]{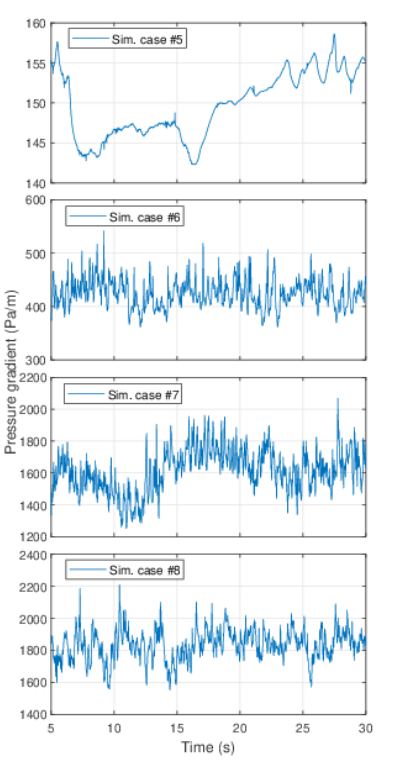}}
\caption{Pressure gradient as a function of time, horizontal eccentric annulus with E=0.5 and 100k cells/m}
\label{fig:PVT_sim_6xxx}
\end{center}
\end{figure}
\begin{figure}[H]
\begin{center}
{\includegraphics[width = 5.5cm]{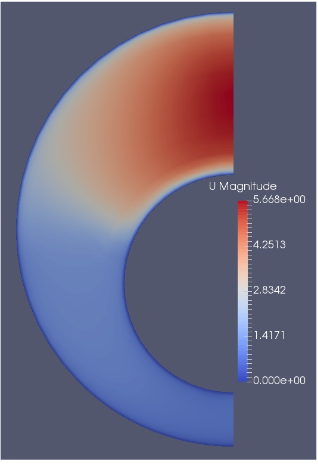}}
\caption{Velocity field of simulation case \#6, horizontal eccentric annulus (E=0.5)}
\label{fig:Usnap}
\end{center}
\end{figure}

The pressure gradient solutions for cases 5 and 6 drastically overshoot the experimental results.  However, the tendency is less surprising when we also account for the change in the central pipe location from E=1.0 to E=0.5. In short, this particular eccentricity change should result in a pressure gradient increase due to the increase in friction factor caused by the development of flow within the narrow gap of the annulus. The narrow gap reduces the velocity of the flow within the region, as shown in Fig. \ref{fig:Usnap}; however, it is still significantly faster than for E=1.0, which contributes to an increased pressure gradient.

The velocity field within the gap is somewhat stagnant compared to the free-flowing liquid within the central region of the annulus. The combination of the stagnant region and the pressure gradient computation method results in an increased pressure gradient when compared to a case with E=1.0. This effect occurs because the slowed down region is a significant contributor to the averaged field, which is accounted for when the simulation determines the pressure gradient required to maintain the mixture velocity.  Physically speaking, there is also an increased friction factor at E=0.5 compared to E=1.0, as stated by \cite{Caetano}, the friction factor is always lower in a more eccentric annulus.

The slug cases are closer to the experimental results, although by no means a perfect match. What we can determine through these case presentations is limited with regards to the pressure gradient solutions themselves without also accounting for the flow regime. It is highly likely, however, that the leading cause of the increased pressure gradients is the different eccentricities of the simulated (E=0.5) and experimental (E=1.0). In addition to the mentioned stagnant flow, in some cases, the partly eccentric configuration results in a partial submersion of the interior pipe, which significantly alters the flow conditions compared to an entirely submerged inner pipe in a fully eccentric configuration. 

However, we can say something about possible reasons why the slug cases are better representations of the pressure solutions than the wave solutions. The slug case pressure distributions experience a higher degree of dependency of the slug structures with regards to pressure gradient. As the proto-slugs form, they create a large fluid structure. The slug, which has a higher mixture density and viscosity, requires more force to push through the domain. As such, when there are slugs present, the pressure gradient is dominated by density and turbulent forces compared to the low velocity cases, where perhaps viscosity and laminar friction are prevalent.

\begin{figure}[H]
\begin{center}
{\includegraphics[width = 8.0cm]{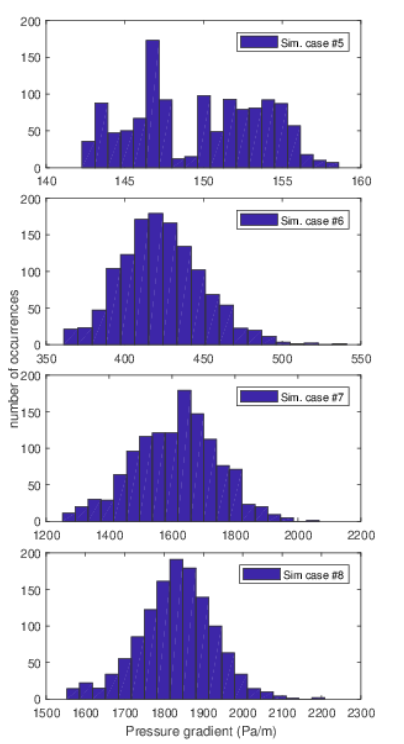}}
\caption{Pressure gradient histogram, horizontal eccentric annulus with E=0.5 and 100k cells/m.}
\label{fig:pa_hist_sim_6xxx}
\end{center}
\end{figure}

\begin{table}[H]
\caption{Pressure gradient summary,horizontal eccentric annulus with E=0.5 and 100k cells/m.}
\begin{center}
\begin{tabular}{|c|c|c|c|}
\hline
Sim, \# &  5\% (Pa/m) &  mean (Pa/m) & 95 \% (Pa/m) \ \\
\hline\hline
5 & 143.45 & 149.69 & 155.67   \\
\hline
6  &  384.25 & 423.59 & 467.41 \\
\hline
7 & 1377.30 & 1612.39 & 1818.22  \\
\hline
8  & 1662.94 & 1831.31 & 1981.58 \\
\hline
\end{tabular}
\label{tab:pressure_info_sim_6xxx}
\end{center}
\end{table}

Comparing the two sets of simulation cases with the same mesh density  (Tabs. \ref{tab:pressure_info_ec} \& \ref{tab:pressure_info_sim_6xxx}), the higher liquid holdup results in a significantly higher pressure gradient for all cases of the same mixture velocity, sometimes dramatically, as shown by the difference in simulation case 2 and 6. Simultaneously, comparing the simulation to the experiments (Tabs. \ref{tab:pressure_info_sim_6xxx} \& \ref{tab:6series:pa:exp}), the relative differences are quite large. We attribute the majority of the change to two aspects, the coarse mesh and the gap left between the inner and outer cylinder, which increases the friction factor. However, even if we account for mesh associated errors, the pressure gradients are drastically higher than for the fully eccentric experiments. 

\begin{table}[H]
\caption{Relative difference of simulations with respect to eccentric experiments for horizontal simulations at E=0.5 and 100k cells/m.}
\begin{center}
\begin{tabular}{|c|c|c|c|}
\hline
Sim \# &  5\% (\%) &  mean (\%) & 95 \% (\%) \ \\
\hline\hline
5 & 838.2 & 312.7 & 58.5   \\
\hline

6 & 413.0 & 311.5 & 241.3  \\
\hline
7 & 92.0 & 67.8  & 36.4  \\
\hline
8 & 61.2 & 51.8 & 41.03   \\
\hline
\end{tabular}
\label{tab:rel_error_fully_ecc}
\end{center}
\end{table}

\begin{figure}[H]
\begin{center}
{\includegraphics[width = 8.0cm]{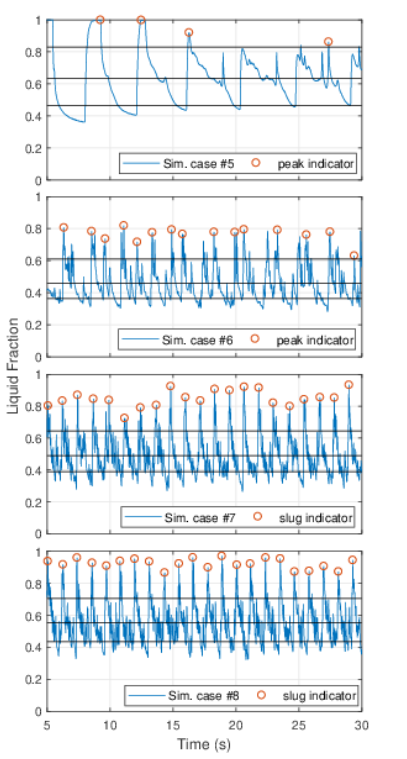}}
\caption{Holdup as a function of time, horizontal eccentric annulus with E=0.5 and 100k cells/m}
\label{fig:HUVT_sim_6xxx}
\end{center}
\end{figure}

As shown in Tabs. \ref{tab:rel_error_fully_ecc} \& \ref{tab:rel_error_100k_case1_4}, the differences are higher when we compare with experimental data at E=1.0, which we expected. We believe the cause of the substantial differences in Sim. 5 is a persistent slug formed at startup, as seen in Fig. \ref{fig:HUVT_sim_6xxx}.

Similarly to simulations compared with E=0, the slug cases perform significantly better; however, we notice the expected pressure gradient increase when comparing simulations of a partly eccentric annulus with fully eccentric experiments.

As can be seen in Fig. \ref{fig:HUVT_sim_6xxx}, simulations 5 to 8 undergo a behavioral change for each case compared to cases 1-4. If we look past the average holdup increase, which is stipulated by the initial conditions, we can also notice that there are some fundamental changes to the holdup pattern. Case 2 has shifted from high frequency small waves (Fig. \ref{fig:HUVT_3series}) to large periodic waves (Fig. \ref{fig:HUVT_sim_6xxx}, Sim case \#6), or potentially proto-slug. The drastically altered pressure gradient solution suggests the simulation is in a different flow state compared to the experiment, which is an intriguing result of the eccentricity.  

We also notice a similar change in the behavior of case 5. Past the initial transient, structures emerge that imply infrequent large waves or proto-slugs followed by a region of smaller waves as the holdup drops off. If wavy, this behavior is consistent with that seen during the experiments and signifies a drastic change to the flow regime when compared to the lower holdup case shown in Fig. \ref{fig:HUVT_3series}. 

It is important to note that the initial transient slug structure observed during simulation case 6 is likely formed by poor initial conditions, which results in a prolonged slug. For most cases, the resolution of the transient occurs during the first 5 s of the simulation. However, for this particular case with a very low average mixture velocity ( 1.2 m/s), the transient persists for nearly 15 s, after which the holdup behavior is closer to the experiment. Even so, the significant pressure gradient errors combined with the holdup pattern indicate that both simulation cases 5 and 6 behave fundamentally different compared to their experiment counterparts.

The slug frequencies of cases 7 and 8 are 0.84 and 0.88 Hz respectively. These values are consistent with the simulation cases 3 and 4. However, particularly for case 8 it is a marked decrease compared to the experiment ehich was 1.64 Hz.

\subsubsection{4 degree inclined simulation}
We simulate the inclined eccentric case with the modified version of interFoam described briefly Sec. \ref{sec:interFoam}  and a 150k cells/m mesh within a 5 m domain. When using periodic boundary conditions, without this modification, the solver may be unable to converge regardless of the number of allowed iterations. The convergence problem for the unmodified interFoam becomes more prevalent with increasing inclination; however, for horizontal cases, the problem is non-existent. 

\begin{figure}[H]
\begin{center}
{\includegraphics[width = 8.0cm]{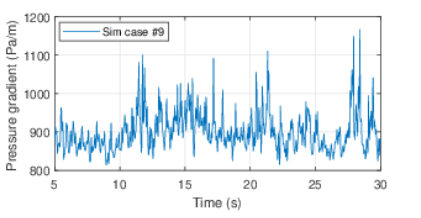}}
\caption{Pressure as a function of time, inclined eccentric simulation with E=0.5}
\label{fig:PVT_sim_7049}
\end{center}
\end{figure}
The pressure gradient of the $4^{o}$ inclined simulation is several hundred Pa/m  higher than the experiment case. This behavior is consistent with the other simulation cases, which are run at $E=0.5$ while their experiment counterparts are at E=1.0. Tab. \ref{tab:exp_7049:pa} shows that for the experiments, the deviation from mean to the 5 and 95 \% thresholds is roughly 35-45 \% of the mean. Comparatively, the simulations maximum and minimum thresholds are within ~10\% of the mean pressure gradient value. Even when accounting for the increased mean pressure gradient, the numerical difference is still significantly smaller for the simulated case.

\begin{figure}[H]
\begin{center}
{\includegraphics[width = 8.0cm]{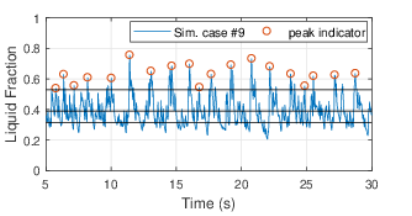}}
\caption{ Holdup as a function of time, inclined eccentric simulation with E=0.5}
\label{fig:HU_sim_7049}
\end{center}
\end{figure}

 \begin{table}[H]
\caption{Pressure gradient summary, inclined eccentric simulation (E=0.5).}
\begin{center}
\begin{tabular}{|c|c|c|c|}
\hline
Sim \# & 5\% (Pa/m) & Mean (Pa/m) & 95\% (Pa/m) \\
\hline\hline
9 & 839.2 & 899.1 & 989.4  \\
\hline
\end{tabular}
\label{tab:sim_7049:pa}
\end{center}
\end{table}
Compared to the fully eccentric experiments (Tab \ref{tab:exp_7049:pa}), we notice that there has been a significant increase in the simulated pressure gradient (Tab \ref{tab:sim_7049:pa}). The minimum is increased by  118 \% while the mean and maximum are increased by 49 and 12 \% respectively. Keep in mind that we expect some increase in the pressure gradient because of the coarseness of the mesh. However, we believe the majority of this discrepancy is caused by  the eccentricity  of the simulations (E=0.5) compared to the experiments (E=1.0) and subsequent a relatively significant change in the flow regime (Figs. \ref{fig:HU_exp_7049} \& \ref{fig:HU_sim_7049}).

\begin{figure}[H]
\begin{center}
{\includegraphics[width = 8.0cm]{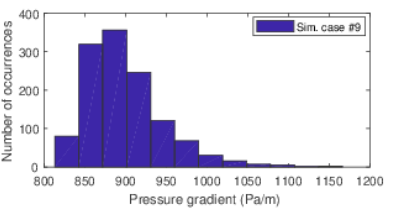}}
\caption{ Pressure  gradient  histogram,  inclined  eccentric  simulation (E=0.5)}
\label{fig:Pa_hist_sim_7049}
\end{center}
\end{figure}
The histogram (Fig. \ref{fig:Pa_hist_sim_7049}) provides a second visual representation of the behaviour of the pressure gradient. As noted when discussing the threshold data (Tab. \ref{tab:sim_7049:pa}), when we cross-reference with the experiment histogram (Fig. \ref{fig:Pa_hist_exp_7049}), we notice that the bulk of the data is more concentrated about the mean and also that the distribution is altered. While for the experiment the distribution is closely resembling a Gaussian distribution, the simulation on the other hand is a right skewed distribution, with a few high pressure gradient data points extending the right tail to 1150 Pa/m.

The flow pattern (Fig. \ref{fig:HU_sim_7049}) appears to be representative of the experiments (Fig. \ref{fig:HU_exp_7049}); however, the peak holdup is slightly higher in the simulations.  The simulation typically peaks at a liquid holdup of between 0.6 and 0.7, while the experiment rarely exceeds 0.6. The frequency of the holdup peaks is 0.76 Hz, which is 0.2 Hz higher than the experimental measurement. 
During the simulation, the flow regime which develops is reminiscent of the wavy flow identified from the experimental data  (Fig. \ref{fig:HU_exp_7049}).

In summary, the holdup pattern and peak frequency are in good agreement with the experiments however we experience an elevated pressure gradient, in particular of the minimum pressure gradient. The pressure gradient increase is slightly larger than what is expected from the difference in eccentricity of the two cases. However, we typically observe a moderate pressure gradient increase at this mesh density.

\begin{figure}[H]
\begin{center}
{\includegraphics[width = 8.0cm]{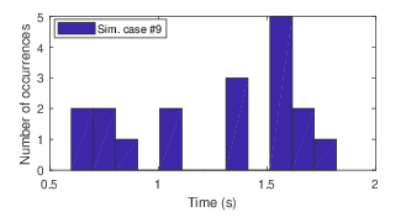}}
\caption{Proto-slug time-wise separation,  inclined  eccentric  simulation with E=0.5}
\label{fig:hz_sim_7049}
\end{center}
\end{figure}
The average proto-slug frequency is 0.76 Hz, with the majority of instances separated in two intervals between 0.5 to 1.1 s and 1.3 to 1.8 s. The distribution of simulated wave (Fig. \ref{fig:hz_sim_7049}) events coincides with the large wave distribution observed in the experiment (Fig. \ref{fig:Hz_Exp_7049})

\section{Conclusions}
Two-phase flow simulations were run in an eccentric annulus using OpenFoam and periodic boundary conditions. The simulations consist of 9 individual cases, of which 8 are horizontal, and 1 is inclined at $4^{\circ}$. The annulus eccentricity was 0.5 for all cases.

For the horizontal domain, the simulations resulted in two distinct flow regimes. We compared the simulations with experimental data collected in both a fully eccentric and fully concentric configuration in order to study the effect of altering the eccentricity. The flow regimes of the experiments were wavy flow or slug flow; the simulations represented these well in all but two cases (case 5 \& 6), where we arguably observe an altered flow state as an effect of the different eccentricity. We show that altering the liquid fraction and location of the inner pipe has an effect on the flow, such as altering the wave and slug frequency, increasing the pressure gradient, and in simulation cases 5 and 6 transitioning the flow to a different flow state. 

The horizontal cases simulated were split into cases referred to as 1-4 and 5-8. These two sets are  cases with the same mixture velocity; however, different fluid fractions. In experiment, they result in similar flow regimes with some minor differences with regard to wave and slug frequency (\ref{tab:3series:hz:exp} \& \ref{tab:6series:hz:exp}) as well as pressure gradient (\ref{tab:3series:pa:exp} \& \ref{tab:6series:pa:exp}). 

As noted in previous works \citep{Friedemann}, the slug structures noted in the simulations resemble proto-slugs, a precursor to a slug, which tends to leave a gap near the upper wall filled with gas. The cause of the void is thought to be an inability to resolve minor bubbles, which would typically permeate through the liquid layer of the slug. When the VOF type solver interFoam is not able to resolve these, the gas bubbles tend to coalesce and rise to the top in order to follow conservation laws.

By comparing the experimental data with the simulations, we can establish that we can simulate both wavy and flow resembling slug structures within an eccentric domain. The cases conform to the experimental data within reasonable tolerance with regards to holdup pattern.  However, we notice a significant discrepancy of the pressure gradient results due to a change of eccentricity.

We did not expect that the simulations would be perfect replications for several reasons. One is that the simulations utilize periodic boundary conditions, which impose some restrictions on the results. For example, the domain length can affect the number of slugs present and thereby alter the slug frequency. The geometry simulated has an eccentricity of 0.5, while the experiments are at E=0 and E=1.0. We purposefully adjusted the eccentricity of the simulation cases to analyze the impact of eccentricity on the pressure gradient and flow regime. 

The most significant deviations from experiments occur when comparing results with a completely eccentric experiment (E=1.0).  For one wavy flow case, we observed a mean pressure gradient increase of as much as 303 \% (Tab. \ref{tab:rel_error_fully_ecc}). The simulation slug case results are generally closer to the experimental case results, and the deviation caused by the change of eccentricity is within 65 \% for the cases where we compare to an eccentric experiment, and 20\% when comparing to a concentric experiment (Tabs. \ref{tab:rel_error_100k_case1_4} \& \ref{tab:rel_error_200k_case1_4}). Based on previous work, the mesh densities studied here typically overestimate the pressure gradients by a modest amount. While a finer mesh reduces the mesh associated errors, the simulation time is sufficiently prohibitive that we opted for the coarser mesh. 

Through these studies, we have learned about the effect of eccentricity on the flow regime and pressure gradient. We notice that the pressure gradient of the slug flow cases behave as expected, the reduction of the pressure gradient corresponds with an increased eccentricity, and vice versa. We also note that for the low velocity cases, the simulations tend to overestimate the pressure gradient significantly; however, they too conform to the expected effect of eccentricity as we note a smaller overshoot when increasing the eccentricity. The flow regime shows a lesser dependence on eccentricity for the slug cases than the wavy flow cases. In particular, for simulation case \# 2, we notice a significantly increased wave frequency when compared to the corresponding experiment. Through the cases studied, we establish that both the flow regime and pressure gradient is altered through the location of the interior pipe. While we do not observe a complete change of flow regime, we do observe mostly minor changes in the wave and slug frequency accompanied by significant changes to the pressure gradient.

\section*{References}

\bibliography{mybibfile}

\end{document}